\documentclass[revtex, amsmath, amssymb,superscriptaddress, reprint]{revtex4-2}

\usepackage[utf8]{inputenc}
\usepackage{graphicx}
\usepackage{dcolumn}
\usepackage{bm}
\usepackage[utf8]{inputenc}
\usepackage[T1]{fontenc}
\usepackage{mathptmx}
\usepackage{gensymb}
\usepackage{upgreek}
\usepackage{xcolor}
\usepackage{textcomp}
\usepackage{hyperref}
\usepackage{mhchem}
\usepackage{lineno}
\usepackage{mathtools,amssymb}
\usepackage{siunitx}

\hypersetup{colorlinks=true,linkcolor=red,citecolor=blue,filecolor=magenta,urlcolor=magenta}

\usepackage{titlesec}
\titleformat{\section}[hang]{\small\bfseries\sffamily}{\thesection.}{0.5em}{\MakeUppercase}
\titlespacing{\section}{0pc}{1pc}{0.2pc}

\begin{document}
\title{High cooperativity coupling of rare earth spins with planar superconducting resonator}

\author{Sihao Wang}
\affiliation{Department of Electrical Engineering, Yale University, New Haven, CT 06511, USA}
\author{Likai Yang}
\affiliation{Department of Electrical Engineering, Yale University, New Haven, CT 06511, USA}
\author{Rufus L. Cone}
\affiliation{Department of Physics, Montana State University, Bozeman, MT 59717, USA}
\author{Charles W. Thiel}
\affiliation{Department of Physics, Montana State University, Bozeman, MT 59717, USA}
\author{Hong X. Tang}
\email{hong.tang@yale.edu}
\affiliation{Department of Electrical Engineering, Yale University, New Haven, CT 06511, USA}


\begin{abstract}
Interfacing superconducting microwave resonator with rare earth doped crystals presents a promising hybrid quantum system for applications including spin-assisted transducers and memories. The coupling strength between spins of rare earth ions and the microwave photons is characterized by the cooperativity. Here we report an ultra-high cooperativity $C \sim$ 650 between rare earth spins and a planar superconducting microwave resonator that features a highly uniform magnetic field for harnessing the strong anisotropic coupling strength of erbium doped yttrium orthosilicate. This cooperativity rivals that from a bulk dielectric resonator and paves a path for efficiently coupling of spins with microwave photons on an integrated platform. 

\end{abstract}
\maketitle

Rare earth ions (REI) have been an attractive candidate for quantum information processing thanks to their favorable optical and spin properties. Stable optical transition and high fluorescence quantum efficiency allow spin control and detection through optical photons \cite{zhong2017nanophotonic, chen2020parallel}. Extremely long coherence time from electronic and nuclear spins makes REIs an ideal choice for quantum information storage \cite{zhong2015optically, ma2021one}. Their narrow homogeneous linewidth further presents opportunities for spectral multiplexing in a broader inhomogeneous linewidth of the spin ensemble \cite{babbitt2014spectral}. Among rare earth elements, erbium (Er) emerges as a popular choice because of its optical transitions in the telecommunication band. This avoids the need for frequency conversion and facilitates integration with existing fiber optic technology.

Common choices of host crystals for REIs include yttrium orthosilicate (\ce{Y2SiO5}), yttrium orthovanadate (\ce{YVO4}), and lithium niobate (\ce{LiNbO3}). Yttrium orthosilicate (YSO) gains its popularity thanks to its weak environmental nuclear magnetic moments. 
Efficient long-lived multimode photonic memories \cite{hedges2010efficient, sabooni2013efficient, laplane2017multimode} have been demonstrated using REI doped YSO, which enables entanglement storage \cite{clausen2011quantum}, quantum teleportation \cite{bussieres2014quantum}, and quantum state transfer from atomic vapor \cite{maring2017photonic}. 

Strong microwave coupling with spins is essential in enabling efficient quantum state manipulation \cite{bienfait2016controlling}, isolating a spin ensemble from the environmental spin bath \cite{albanese2020radiative}, and relaxing the optical coupling requirement for a high efficiency microwave-to-optical transducer \cite{williamson2014magneto}. Due to the large mode size mismatch, and hence the poor mode overlap between resonant spins and the microwave photon field, microwave resonators are often employed to enhance their coupling strengths. These include both bulk three-dimensional (3D) cavities \cite{probst2014three, fernandez2015coherent, chen2018hyperfine, creedon2012four, farr2015evidence} and planar superconducting resonators \cite{wisby2016angle, bushev2011ultralow, staudt2012coupling, probst2013anisotropic, tkalvcec2014strong, dold2019high}. By providing a homogeneous magnetic field (B field) and high spatial mode overlap with the embedded REI-doped crystal, 3D dielectric cavities 
benefit from spin-number enhancement from the REI ensemble and enable an ultra-high cooperativity of $C\sim600$ 
\cite{farr2015evidence}. On the other hand, planar superconducting resonators offer a small device footprint and mode volume, and thus enhanced single spin coupling rates. However, the magnetic field distribution tends to become less uniform with microscale patterned structures, thus reducing the effective modal overlap between spins and the RF magnetic field. Experiments on coupling to rare earth ions have so far yielded a cooperativity of $C\sim30$ \cite{probst2013anisotropic, dold2019high}. Further improvement is required to improve the mode overlap while maintaining a high Q factor. 

In this work, we present an ultra-high cooperativity of $C\sim650$ between Er$^{3+}$ ions and a niobium nitride (NbN) superconducting microwave resonator through designing the parallel nanowire strips for homoegenous B field and utilizing the anisotropic coupling strength in an erbium doped yttrium orthosilicate (Er:YSO) crystal. We use flip chip bonding techniques to integrate the Er:YSO crystal on a NbN superconducting resonator with a Q factor $Q>10^5$. Electron spin resonance (ESR) spectra are measured at different orientations to demonstrate strong coupling and high cooperativity. 

\begin{figure}[h]
\includegraphics[width=0.5\textwidth]{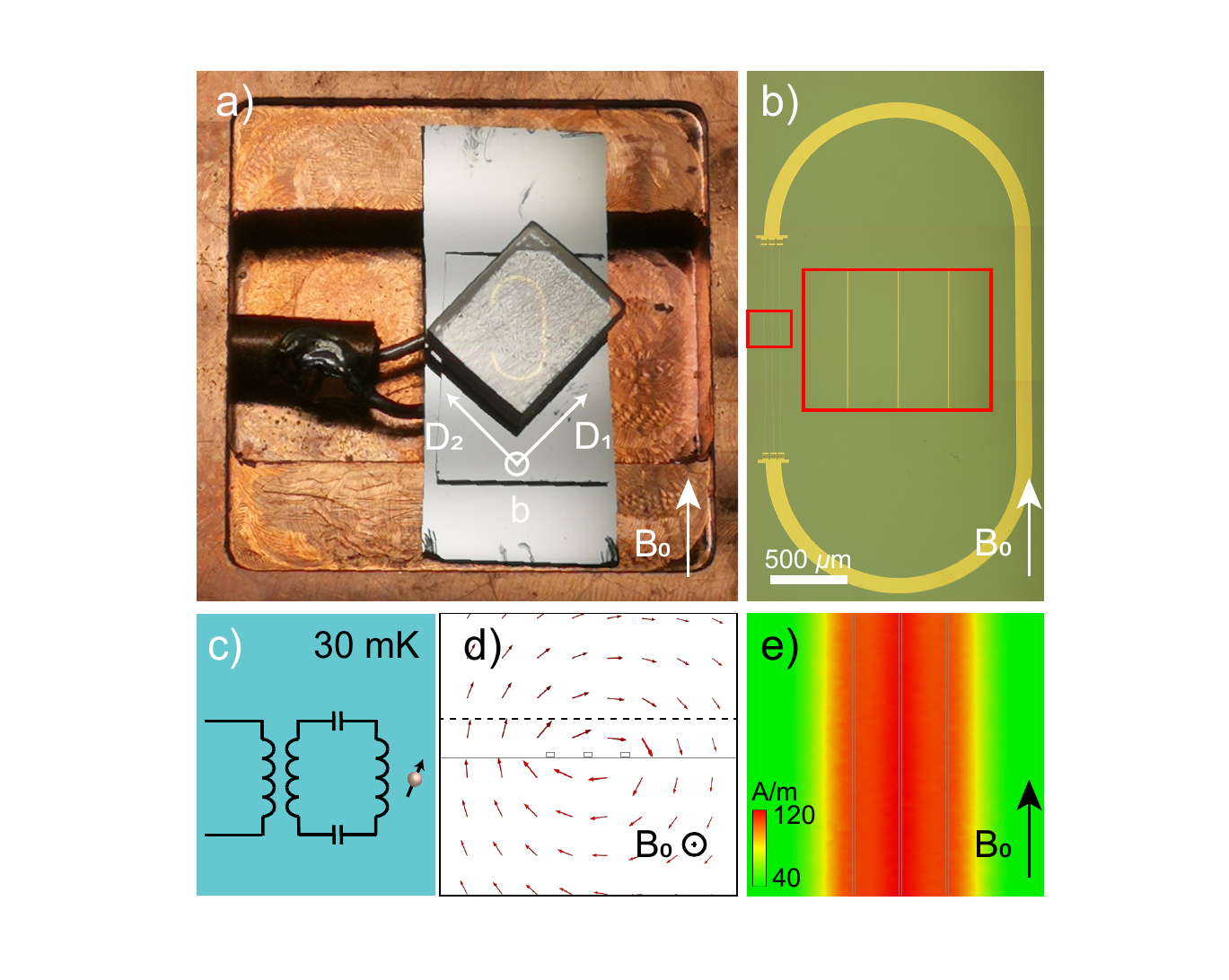}
\caption{\label{fig:1} \textbf{(a)} A photograph of the packaged device. The NbN microwave resonator was fabricated on a high resistivity silicon substrate that sits on a bigger piece of high resistivity silicon carrier in a copper cavity. A hoop antenna was used to couple to the device. The crystal axes are indicated and the applied static magnetic field $B_0$ is oriented parallel to the resonator inductor strips. \textbf{(b)} The optical micrograph of the NbN resonator. A zoom-in image of the narrow inductor strips is shown in the inset. \textbf{(c)} The corresponding circuit model of the device. \textbf{(d)} The cross-sectional arrow plot of the oscillating magnetic field $B_1$ around the inductor strips. The thin strips are exaggerated for viewing purposes. \textbf{(e)} The oscillating magnetic field profile at 35\,$\mu$m above the microwave resonator, indicated by the dashed line in \textbf{(d)}. The uniform magnetic field facilitates a narrow spin linewidth. The narrow inductor strips are exaggerated for illustration purpose.}
\end{figure}

A commercial bulk 10\,ppm natural abundance Er:YSO crystal (Scientific Materials) was cut into a cuboid (5\,mm $\times$ 4\,mm $\times$ 3\,mm) along the three dielectric axes ($D_1$, $D_2$, $b$) and flip chip bonded on a NbN superconducting microwave resonator with the $b$ axis perpendicular to the bonding interface. The thin film of 50\,nm NbN is grown on a high resistivity silicon substrate using atomic layer deposition (ALD), with a sheet resistance $R_{\mathrm{sq}}\sim 45\,\Omega/\mathrm{sq}$. We fabricate microwave resonators by patterning hydrogen silsequioxane (HSQ) on NbN thin film with electron beam lithography (EBL) and etching in reactive ion etching (RIE) with chlorine. The resonator has a shape of a racetrack for strong coupling with the hoop antenna. The width, radius, and straight section length of the racetrack resonator are $100\,\mu$m, $800\,\mu$m and $1500\,\mu$m, respectively. Three parallel $1\,\mu$m-wide strips act as inductors that replace one straight section of the racetrack resonator. These inductor strips are connected to the racetrack resonator through interdigitated (IDT) capacitors. This design allows the generated vortices inside the racetrack resonator to traverse out, hence stabilizing the cavity frequency under a strong external B field. The final flip chip bonded device is housed in a copper cavity box to reduce radiation loss to the environment. Fig.~\ref{fig:1}a) shows a photo of the device inside the copper box. The external static B field $B_0$ is applied parallel to the inductor strips. The oscillating B field $B_1$ from the inductor strips is therefore perpendicular to $B_0$. The use of flip chip bonding permits free orientation of the Er:YSO crystal on the microwave resonator. Fig.~\ref{fig:1}b) is an optical micrograph of the NbN microwave resonator. The narrow inductor strips are magnified in the inset. The corresponding circuit model is shown in Fig.~\ref{fig:1}c), where the microwave resonator is modeled as an LC resonator and the hoop antenna as an inductor. Fig.~\ref{fig:1}d) is the cross-sectional plot of the oscillating B field $B_1$ from the inductor strips. As opposed to the commonly used meander inductors where the current in the adjacent straight sections flows in opposite directions, the current in the three parallel inductor strips flows in the same direction, resulting in a more uniform B field $B_1$ inside the bonded Er:YSO crystal. Fig.~\ref{fig:1}e) is the oscillating B field $B_1$ strength at 35\,$\mu$m above the NbN microwave resonator, indicated by the dashed line in Fig.~\ref{fig:1}d). A homogeneous B field distribution above the inductor strips plays an important role in reducing the spin linewidth.     

The packaged device is loaded in the mixing chamber of a dilution fridge with a base temperature of 10\,mK. The actual device temperature is 30\,mK due to a thermal gradient. A vector network analyzer (VNA) is used to measure the $S_{21}$ response. The microwave resonator has a resonance frequency at 5.7\,GHz at zero applied B field $B_0=0$. The Q factor of the cavity is fitted with a Fano resonance to take into account the background response.  Q $\sim 130000$ is obtained at the input power of -70\,dBm, at which the loss from two level systems is minimized.  

\noindent The Hamiltonian of the coupled system is given as
\begin{eqnarray}
H = \mu_B \mathbf{B} \cdot \mathbf{g} \cdot \mathbf{S} - \mu_n g_n \mathbf{B} \cdot \mathbf{I} + \mathbf{I} \cdot \mathbf{Q} \cdot \mathbf{I} + \mathbf{S} \cdot \mathbf{A} \cdot \mathbf{I},
\label{eq:1}
\end{eqnarray}
where $\mu_B$ ($\mu_n$) is the electron (nuclear) Bohr magnetron, $g_n = -0.1618$ is the nuclear $g$-factor, $\mathbf{g}$ is the anisotropic electron $g$-factor tensor, $\mathbf{Q}$ is the nuclear quadruple tensor and $\mathbf{A}$ is the hyperfine interaction tensor. The first two terms of the spin Hamiltonian are electron and nuclear Zeeman splittings, which are proportional to the applied B field $B_0$. The third and the fourth term are the nuclear quadruple interaction and the hyperfine interaction, independent of $B_0$. Based on the published tensor values \cite{guillot2006hyperfine, sun2008magnetic}, we can calculate the eigen-energy spectra of the Er:YSO crystal at different orientations \cite{stoll2006easyspin}. 

\begin{figure}[htbp!]
\includegraphics[width=0.5\textwidth]{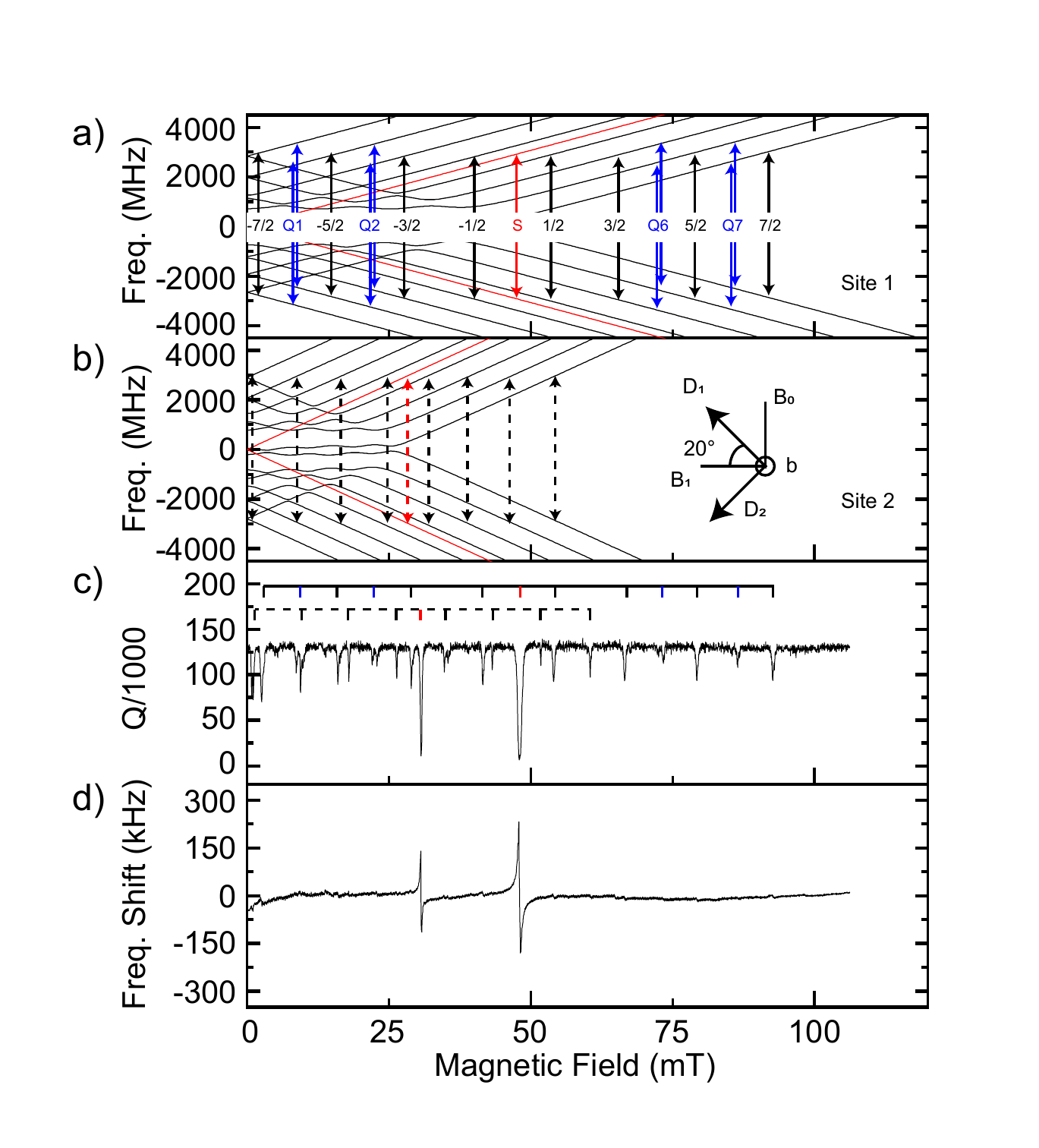}
\caption{\label{fig:2} \textbf{(a)} The simulated Breit-Rabi diagram of Er:YSO for spins residing at site 1 subject to an oscillating magnetic field $B_1$ oriented 20$^{\circ}$ from D$_1$ axis in the D$_1$-D$_2$ plane. The allowed and forbidden transitions at 5.7 GHz are shown in solid arrows with different colors. \textbf{(b)} The corresponding Breit-Rabi diagram 
for spins at site 2. The allowed transitions are shown by dashed arrows with different colors. The inset shows a schematic of the orientation of the YSO crystal with respect to the magnetic fields. \textbf{(c)} The spectrum of the loaded Q factor of the superconducting microwave resonator with the static magnetic field $B_0$. Each dip in the spectrum indicates a level anti-crossing between the spin and microwave modes. Transitions from site 1(2) are indicated with solid(dashed) markers. \textbf{(d)} The frequency shift of the ESR spectrum with the static magnetic field $B_0$. The quadratic dependence of the superconducting microwave resonance with the static magnetic field $B_0$ has been normalized.}
\end{figure}

The crystal structure of Er:YSO is described by space group $C_{2h}^6$, where Er$^{3+}$ ions substitute Y$^{3+}$ ions at two distinct crystallographic sites, both characterized by $C_1$ symmetry. For each crystallographic site, the crystal's $C_2$ rotation symmetry and inversion symmetry give rise to four subclass sites in the crystal with different local orientations. Based on their response to the applied B field, these subclass sites can be further divided into two groups. The group related to the inversion symmetry is magnetically equivalent as it responds to a B field of arbitrary direction identically. The other group related to the $C_2$ symmetry is magnetically inequivalent. These two subclass sites interact with an applied B field differently, except for two special cases, when the B field is either applied along the $b$ axis or in the $D_1$-$D_2$ plane. During the measurement, a static B field $B_0$ is applied parallel to the inductor strips in the $D_1$-$D_2$ plane to minimize the vortex-induced loss. The oscillating B field $B_1$ from the inductors is perpendicular to $B_0$. Fig.~\ref{fig:2}a) and Fig.~\ref{fig:2}b) are the Breit-Rabi diagrams of Er$^{3+}$ ions at site 1 and site 2 when the oscillating B field $B_1$ is $\sim 20^{\circ}$ from the $D_1$ axis. The strong electron spin transitions due to the electron Zeeman term at both sites under the selection rule $\Delta m_s = \pm 1$ are marked by the red arrows. The odd number isotope $^{167}$Er has a nuclear spin of $I=7/2$, resulting in the satellite hyperfine spin state transitions (black arrows) with $\Delta m_i = 0$ and the nuclear quadruple transitions (blue arrows) with $\Delta m_i = \pm 1$. 

The length of the arrows are selected to match the microwave resonance frequency to demonstrate the positions of level anti-crossing. This level anti-crossing is manifested as the coupling of the microwave field to the spins and hence an additional loss channel of the resonator, resulting in a drop of Q factor of the microwave resonance. In the case of perfect alignment, only one level anti-crossing for Er$^{3+}$ ions, and therefore one dip of the Q factor at each site, should be observed when sweeping the B field strength. However, two dips are observed (see Supplementary Information for details), indicating a small misalignment between the applied B field and the $D_1$-$D_2$ plane. By utilizing an additional balance coil with applied B field in the perpendicular direction, we are able to cancel the out-of-plane component of $B_0$. The deviation angle is estimated to be $\sim 1.05^{\circ}$, which is comparable to the expected uncertainty in the crystal orientation. Fig.~\ref{fig:2}c) shows the variation of Q factor with the sweeping B field. Solid (dashed) markers represent the assignment of each level anti-crossing to the spin state transition at each site in Fig.~\ref{fig:2}a) and b). The prominent dips correspond to the strong Zeeman electron spin state transitions and the satellite dips correspond to the transitions with nuclear hyperfine states. Fig.~\ref{fig:2}d) is the corresponding frequency shift of the microwave resonance at level anti-crossing fields. The background quadratic dependence of the microwave resonance frequency on $B_0$ due to the nonlinearity from the kinetic inductance in the narrow inductor strips is normalized to highlight the frequency shift.

\begin{figure}[h]
\includegraphics[width=0.5\textwidth]{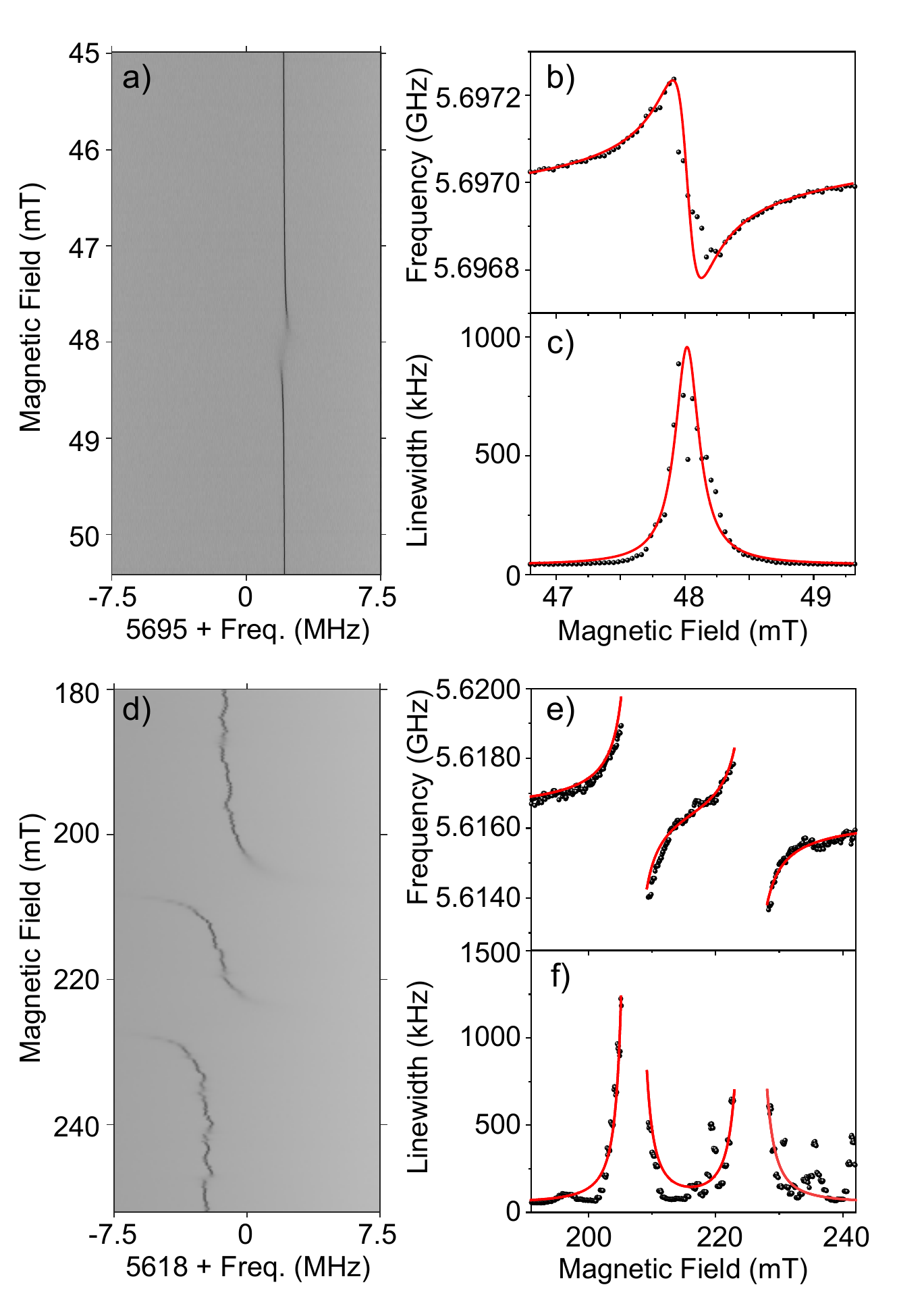}
\caption{\label{fig:3} \textbf{(a-c)} The electron spin level anti-crossing for the oscillating magnetic field $B_1$ oriented 20$^{\circ}$ from the D$_1$ axis in the D$_1$-D$_2$ plane. \textbf{(a)} The S$_{21}$ spectra of the microwave resonance when sweeping the static magnetic field $B_0$ across the level anti-crossing point. \textbf{(b)} The frequency shift as a function of the applied static magnetic field $B_0$ and the fitting. \textbf{(c)} The resonator linewidth as a function of the applied static magnetic field $B_0$ and the fitting. \textbf{(d-f)} The electron spin level anti-crossing for the oscillating magnetic field $B_1$ oriented 112$^{\circ}$ from the D$_1$ axis. The two subclass modes indicate that the magnetic field is slightly off from the D$_1$-D$_2$ plane. \textbf{(d)} The S$_{21}$ spectra of the microwave resonance when sweeping the static magnetic field $B_0$ across the level anti-crossing point. \textbf{(e)} The frequency shift as a function the applied static magnetic field $B_0$ and the fitting. \textbf{(f)} The resonator linewidth as a function of the applied static magnetic field $B_0$ and the fitting.   }
\end{figure}

The collective coupling $\nu$ between the spin ensemble and the microwave resonator is proportional to the effective $g$-factor $g$ by $\nu = g \mu_B \sqrt{n\xi \omega_0 \mu_0/(4 \hbar)} $, where $\mu_B$ is the Bohr magneton, $n\sim 10^{17}$\,cm$^{-3}$ is the concentration of Er$^{3+}$ ions, $\xi \sim 0.25$ is the geometry factor accounted for the mode overlap with spins and the spin coupling with the oscillating field, and $\omega_0$ is the microwave resonance frequency. For an oscillating B field $B_1$ in the $D_1$-$D_2$ plane, the collective coupling strength is maximized when the angle between $B_1$ and the $D_1$ axis is close to $120^{\circ}$ and minimized when the angle is near $30^{\circ}$. Fig.~\ref{fig:3}a) shows the evolution of the microwave resonance with the sweeping B field $B_0$ when $B_1$ is at the angle of $20^{\circ}$ to the crystal axis $D_1$ for Er$^{3+}$ ions at site 1. The level anti-crossing near $B_0=48$\,mT can be extracted. The modified resonator frequency $\omega$ and decay rate $\kappa$ is fitted with a simplified phenomenological model \cite{bushev2011ultralow, dold2019high}
\begin{eqnarray}
\omega &=& \omega_0 - \frac{ \nu^2 \Delta }{\Delta^2 + \gamma_s^2/4},
\label{eq:2}\\
\kappa &=& \kappa_c + \frac{ \nu^2 \gamma_s }{\Delta^2 + \gamma_s^2/4},
\label{eq:3}
\end{eqnarray}
where $\Delta = \omega - \omega_0$ is the detuning, $\gamma_s$ is the spin inhomogeneous linewidth, and $\kappa_c$ is the initial microwave cavity decay rate. Fig.~\ref{fig:3}b) and c) show fitting of the frequency shift and linewidth, yielding the collective coupling strength $\nu = (2\pi) 2.5$\,MHz, the spin ensemble inhomogeneous linewidth $\gamma_s = (2\pi) 26.6$\,MHz, and the cavity linewidth $\kappa_c = (2\pi) 41$\,kHz. The cooperativity is thus $C = 4\nu^2/(\kappa_c \gamma_s) = 22$. 

For a large cooperativity, we rotate the Er:YSO crystal so that the angle between $B_1$ and the $D_1$ axis is 112$^{\circ}$, near 120$^{\circ}$ for maximum $g$. Fig.~\ref{fig:3}d) is a similar plot of the resonance evolution with sweeping $B_0$ when $B_1$ is at 112$^{\circ}$ to the crystal axis $D_1$ for Er$^{3+}$ ions at site 1. At this orientation, the coupling of the spins to the oscillating B field $B_1$ is maximized, but the interaction with the static B field $B_0$ is minimized. The out-of-plane component of $B_0$ is beyond the compensation range of the balance coil. Therefore, no balance coil is applied at 112$^{\circ}$ and the applied $B_0$ is $\sim 1.05^{\circ}$ deviated from the $D_1$-$D_2$ plane. This deviation results in the lifting of the degeneracy of the subclass sites related to the $C_2$ rotation of the crystal lattice.

The resonator frequency and decay rate are fitted with a modified model to account for the double level anti-crossings from the subclass levels.
\begin{eqnarray}
\omega &=& \omega_0 - \frac{\nu_1^2 \Delta}{\Delta^2 + \gamma_1^2/4} - \frac{\nu_2^2 \Delta}{\Delta^2 + \gamma_2^2/4},
\label{eq:4}\\
\kappa &=& \kappa_c + \frac{\nu_1^2 \gamma_1}{\Delta^2 + \gamma_1^2/4} + \frac{\nu_2^2 \gamma_2}{\Delta^2 + \gamma_2^2/4},
\label{eq:5}
\end{eqnarray}
where $\nu_{1(2)}$ and $\gamma_{1(2)}$ are the collective coupling strength and the linewidth for ions at subclass level 1 (2) respectively. Fig.~\ref{fig:3}e) and f) show the fitting of the frequency shift and linewidth, yielding $\nu_1 = (2\pi) 12.5$\,MHz, $\nu_2 = (2\pi) 12.2$\,MHz, $\gamma_1 = (2\pi) 18.5$\,MHz, $\gamma_2 = (2\pi) 18.1$\,MHz and $\kappa_c = (2\pi) 51$\,kHz. The corresponding cooperativities for each subclass are $C_1 = 4\nu_1^2/(\kappa_c \gamma_1) = 650$ and $C_2 = 4\nu_2^2/(\kappa_c \gamma_2) = 634$. 

The cooperativity of $C\sim650$ is one order of magnitude larger than the literature values from cases with planar superconducting microwave resonators, rivaling that from the 3D cavity with high spin number enhancement. Yet, this cooperativity can be further optimized. Our current manual alignment during the flip chip bonding process is not perfect, causing the crystal orientation to deviate from the optimal angle for the $g$-factor. The lifting of degeneracy due to the out-of-plane B field can be suppressed with a larger balance coil, bringing about a two-fold improvement as spins from both subclasses contribute. Further improvement could come from the reduction of vortex-induced microwave loss, hence a smaller $\kappa_c$. These improvements are evidenced by the fitting results from the double level anti-crossings without a balance coil at $20^{\circ}$ (see Supplementary Information for details). Fitting with Eq.~\ref{eq:4} and \ref{eq:5} yields $\nu_1(20^{\circ}) = (2\pi)1.5$\,MHz, $\nu_2(20^{\circ})=(2\pi)1.6$\,MHz, $\gamma_1(20^{\circ}) =(2\pi) 18.7$\,MHz, $\gamma_2(20^{\circ})=(2\pi)18.8$\,MHz and $\kappa_c(20^{\circ})=(2\pi)51$\,kHz. The corresponding cooperativities are $C_1(20^{\circ})=9$ and $C_2(20^{\circ})=11$. Notice that the spin linewidth $\gamma_s$ is higher when the balance coil is applied. This is likely due to the imperfect cancellation of the out-of-plane B field so that the two level anti-crossings do not overlap completely. The flip chip bonding process may still leave a small gap between the crystal and the superconducting resonator, reducing the geometry overlapping factor $\xi$. The theoretical estimation of the coupling strength with zero gap and $g=12$ at $\xi\sim0.25$ is $\nu_{\mathrm{theo}} \sim (2\pi) 28.7$\,MHz, which is more than two times higher than the current extracted value, resulting in a cooperativity of thousands. Direct deposition and fabrication of superconducting resonators on the Er:YSO crystal \cite{dold2019high} is a potential solution for minimizing the gap, but at the expense of the flexibility to orientate the crystal at an arbitrary angle.

In conclusion, we realize high cooperativity coupling between an Er$^{3+}$ spin ensemble and the microwave resonator by producing a more homogeneous B field and exploiting the anisotropic $g$-factor tensor. An ultra-high cooperativity of $C=650$ is achievable, which is comparable to that from large 3D resonators with high spin number enhancement. This result marks an important step towards a high efficiency microwave-to-optical transducer, an essential component in building a coherent link between a superconducting quantum circuit and optical quantum network. The strong coupling also paves the way for spin manipulation and detection such as cooling of spin ensembles and high sensitivity spin detection. An efficient state transfer and mapping in and out of the spin states with strong coupling is highly desired in a quantum memory device. 

\section{Acknowledgement}
This work is supported by Department of Energy, Office of Basic Energy Sciences, Division of Materials Sciences and Engineering under Grant DE-SC0019406. The authors would like to thank Dr. Yong Sun, Sean Rinehart, Kelly Woods, and Dr. Michael Rooks for their assistance provided in the device fabrication. The fabrication of the devices was done at the Yale School of Engineering \& Applied Science (SEAS) Cleanroom and the Yale Institute for Nanoscience and Quantum Engineering (YINQE).\\
\\
\noindent\textbf{REFERENCES}

\bibliography{References}

\begin{thebibliography}{28}%
\makeatletter
\providecommand \@ifxundefined [1]{%
 \@ifx{#1\undefined}
}%
\providecommand \@ifnum [1]{%
 \ifnum #1\expandafter \@firstoftwo
 \else \expandafter \@secondoftwo
 \fi
}%
\providecommand \@ifx [1]{%
 \ifx #1\expandafter \@firstoftwo
 \else \expandafter \@secondoftwo
 \fi
}%
\providecommand \natexlab [1]{#1}%
\providecommand \enquote  [1]{``#1''}%
\providecommand \bibnamefont  [1]{#1}%
\providecommand \bibfnamefont [1]{#1}%
\providecommand \citenamefont [1]{#1}%
\providecommand \href@noop [0]{\@secondoftwo}%
\providecommand \href [0]{\begingroup \@sanitize@url \@href}%
\providecommand \@href[1]{\@@startlink{#1}\@@href}%
\providecommand \@@href[1]{\endgroup#1\@@endlink}%
\providecommand \@sanitize@url [0]{\catcode `\\12\catcode `\$12\catcode
  `\&12\catcode `\#12\catcode `\^12\catcode `\_12\catcode `\%12\relax}%
\providecommand \@@startlink[1]{}%
\providecommand \@@endlink[0]{}%
\providecommand \url  [0]{\begingroup\@sanitize@url \@url }%
\providecommand \@url [1]{\endgroup\@href {#1}{\urlprefix }}%
\providecommand \urlprefix  [0]{URL }%
\providecommand \Eprint [0]{\href }%
\providecommand \doibase [0]{https://doi.org/}%
\providecommand \selectlanguage [0]{\@gobble}%
\providecommand \bibinfo  [0]{\@secondoftwo}%
\providecommand \bibfield  [0]{\@secondoftwo}%
\providecommand \translation [1]{[#1]}%
\providecommand \BibitemOpen [0]{}%
\providecommand \bibitemStop [0]{}%
\providecommand \bibitemNoStop [0]{.\EOS\space}%
\providecommand \EOS [0]{\spacefactor3000\relax}%
\providecommand \BibitemShut  [1]{\csname bibitem#1\endcsname}%
\let\auto@bib@innerbib\@empty
\bibitem [{\citenamefont {Zhong}\ \emph {et~al.}(2017)\citenamefont {Zhong},
  \citenamefont {Kindem}, \citenamefont {Bartholomew}, \citenamefont {Rochman},
  \citenamefont {Craiciu}, \citenamefont {Miyazono}, \citenamefont
  {Bettinelli}, \citenamefont {Cavalli}, \citenamefont {Verma}, \citenamefont
  {Nam} \emph {et~al.}}]{zhong2017nanophotonic}%
  \BibitemOpen
  \bibfield  {author} {\bibinfo {author} {\bibfnamefont {T.}~\bibnamefont
  {Zhong}}, \bibinfo {author} {\bibfnamefont {J.~M.}\ \bibnamefont {Kindem}},
  \bibinfo {author} {\bibfnamefont {J.~G.}\ \bibnamefont {Bartholomew}},
  \bibinfo {author} {\bibfnamefont {J.}~\bibnamefont {Rochman}}, \bibinfo
  {author} {\bibfnamefont {I.}~\bibnamefont {Craiciu}}, \bibinfo {author}
  {\bibfnamefont {E.}~\bibnamefont {Miyazono}}, \bibinfo {author}
  {\bibfnamefont {M.}~\bibnamefont {Bettinelli}}, \bibinfo {author}
  {\bibfnamefont {E.}~\bibnamefont {Cavalli}}, \bibinfo {author} {\bibfnamefont
  {V.}~\bibnamefont {Verma}}, \bibinfo {author} {\bibfnamefont {S.~W.}\
  \bibnamefont {Nam}}, \emph {et~al.},\ }\bibfield  {title} {\bibinfo {title}
  {Nanophotonic rare-earth quantum memory with optically controlled
  retrieval},\ }\href@noop {} {\bibfield  {journal} {\bibinfo  {journal}
  {Science}\ }\textbf {\bibinfo {volume} {357}},\ \bibinfo {pages} {1392}
  (\bibinfo {year} {2017})}\BibitemShut {NoStop}%
\bibitem [{\citenamefont {Chen}\ \emph {et~al.}(2020)\citenamefont {Chen},
  \citenamefont {Raha}, \citenamefont {Phenicie}, \citenamefont {Ourari},\ and\
  \citenamefont {Thompson}}]{chen2020parallel}%
  \BibitemOpen
  \bibfield  {author} {\bibinfo {author} {\bibfnamefont {S.}~\bibnamefont
  {Chen}}, \bibinfo {author} {\bibfnamefont {M.}~\bibnamefont {Raha}}, \bibinfo
  {author} {\bibfnamefont {C.~M.}\ \bibnamefont {Phenicie}}, \bibinfo {author}
  {\bibfnamefont {S.}~\bibnamefont {Ourari}},\ and\ \bibinfo {author}
  {\bibfnamefont {J.~D.}\ \bibnamefont {Thompson}},\ }\bibfield  {title}
  {\bibinfo {title} {Parallel single-shot measurement and coherent control of
  solid-state spins below the diffraction limit},\ }\href@noop {} {\bibfield
  {journal} {\bibinfo  {journal} {Science}\ }\textbf {\bibinfo {volume}
  {370}},\ \bibinfo {pages} {592} (\bibinfo {year} {2020})}\BibitemShut
  {NoStop}%
\bibitem [{\citenamefont {Zhong}\ \emph {et~al.}(2015)\citenamefont {Zhong},
  \citenamefont {Hedges}, \citenamefont {Ahlefeldt}, \citenamefont
  {Bartholomew}, \citenamefont {Beavan}, \citenamefont {Wittig}, \citenamefont
  {Longdell},\ and\ \citenamefont {Sellars}}]{zhong2015optically}%
  \BibitemOpen
  \bibfield  {author} {\bibinfo {author} {\bibfnamefont {M.}~\bibnamefont
  {Zhong}}, \bibinfo {author} {\bibfnamefont {M.~P.}\ \bibnamefont {Hedges}},
  \bibinfo {author} {\bibfnamefont {R.~L.}\ \bibnamefont {Ahlefeldt}}, \bibinfo
  {author} {\bibfnamefont {J.~G.}\ \bibnamefont {Bartholomew}}, \bibinfo
  {author} {\bibfnamefont {S.~E.}\ \bibnamefont {Beavan}}, \bibinfo {author}
  {\bibfnamefont {S.~M.}\ \bibnamefont {Wittig}}, \bibinfo {author}
  {\bibfnamefont {J.~J.}\ \bibnamefont {Longdell}},\ and\ \bibinfo {author}
  {\bibfnamefont {M.~J.}\ \bibnamefont {Sellars}},\ }\bibfield  {title}
  {\bibinfo {title} {Optically addressable nuclear spins in a solid with a
  six-hour coherence time},\ }\href@noop {} {\bibfield  {journal} {\bibinfo
  {journal} {Nature}\ }\textbf {\bibinfo {volume} {517}},\ \bibinfo {pages}
  {177} (\bibinfo {year} {2015})}\BibitemShut {NoStop}%
\bibitem [{\citenamefont {Ma}\ \emph {et~al.}(2021)\citenamefont {Ma},
  \citenamefont {Ma}, \citenamefont {Zhou}, \citenamefont {Li},\ and\
  \citenamefont {Guo}}]{ma2021one}%
  \BibitemOpen
  \bibfield  {author} {\bibinfo {author} {\bibfnamefont {Y.}~\bibnamefont
  {Ma}}, \bibinfo {author} {\bibfnamefont {Y.-Z.}\ \bibnamefont {Ma}}, \bibinfo
  {author} {\bibfnamefont {Z.-Q.}\ \bibnamefont {Zhou}}, \bibinfo {author}
  {\bibfnamefont {C.-F.}\ \bibnamefont {Li}},\ and\ \bibinfo {author}
  {\bibfnamefont {G.-C.}\ \bibnamefont {Guo}},\ }\bibfield  {title} {\bibinfo
  {title} {One-hour coherent optical storage in an atomic frequency comb
  memory},\ }\href@noop {} {\bibfield  {journal} {\bibinfo  {journal} {Nature
  communications}\ }\textbf {\bibinfo {volume} {12}},\ \bibinfo {pages} {1}
  (\bibinfo {year} {2021})}\BibitemShut {NoStop}%
\bibitem [{\citenamefont {Babbitt}\ \emph {et~al.}(2014)\citenamefont
  {Babbitt}, \citenamefont {Barber}, \citenamefont {Bekker}, \citenamefont
  {Chase}, \citenamefont {Harrington}, \citenamefont {Merkel}, \citenamefont
  {Mohan}, \citenamefont {Sharpe}, \citenamefont {Stiffler}, \citenamefont
  {Traxinger} \emph {et~al.}}]{babbitt2014spectral}%
  \BibitemOpen
  \bibfield  {author} {\bibinfo {author} {\bibfnamefont {W.~R.}\ \bibnamefont
  {Babbitt}}, \bibinfo {author} {\bibfnamefont {Z.~W.}\ \bibnamefont {Barber}},
  \bibinfo {author} {\bibfnamefont {S.~H.}\ \bibnamefont {Bekker}}, \bibinfo
  {author} {\bibfnamefont {M.~D.}\ \bibnamefont {Chase}}, \bibinfo {author}
  {\bibfnamefont {C.}~\bibnamefont {Harrington}}, \bibinfo {author}
  {\bibfnamefont {K.~D.}\ \bibnamefont {Merkel}}, \bibinfo {author}
  {\bibfnamefont {R.~K.}\ \bibnamefont {Mohan}}, \bibinfo {author}
  {\bibfnamefont {T.}~\bibnamefont {Sharpe}}, \bibinfo {author} {\bibfnamefont
  {C.~R.}\ \bibnamefont {Stiffler}}, \bibinfo {author} {\bibfnamefont {A.~S.}\
  \bibnamefont {Traxinger}}, \emph {et~al.},\ }\bibfield  {title} {\bibinfo
  {title} {From spectral holeburning memory to spatial-spectral microwave
  signal processing},\ }\href@noop {} {\bibfield  {journal} {\bibinfo
  {journal} {Laser Physics}\ }\textbf {\bibinfo {volume} {24}},\ \bibinfo
  {pages} {094002} (\bibinfo {year} {2014})}\BibitemShut {NoStop}%
\bibitem [{\citenamefont {Hedges}\ \emph {et~al.}(2010)\citenamefont {Hedges},
  \citenamefont {Longdell}, \citenamefont {Li},\ and\ \citenamefont
  {Sellars}}]{hedges2010efficient}%
  \BibitemOpen
  \bibfield  {author} {\bibinfo {author} {\bibfnamefont {M.~P.}\ \bibnamefont
  {Hedges}}, \bibinfo {author} {\bibfnamefont {J.~J.}\ \bibnamefont
  {Longdell}}, \bibinfo {author} {\bibfnamefont {Y.}~\bibnamefont {Li}},\ and\
  \bibinfo {author} {\bibfnamefont {M.~J.}\ \bibnamefont {Sellars}},\
  }\bibfield  {title} {\bibinfo {title} {Efficient quantum memory for light},\
  }\href@noop {} {\bibfield  {journal} {\bibinfo  {journal} {Nature}\ }\textbf
  {\bibinfo {volume} {465}},\ \bibinfo {pages} {1052} (\bibinfo {year}
  {2010})}\BibitemShut {NoStop}%
\bibitem [{\citenamefont {Sabooni}\ \emph {et~al.}(2013)\citenamefont
  {Sabooni}, \citenamefont {Li}, \citenamefont {Kr{\"o}ll},\ and\ \citenamefont
  {Rippe}}]{sabooni2013efficient}%
  \BibitemOpen
  \bibfield  {author} {\bibinfo {author} {\bibfnamefont {M.}~\bibnamefont
  {Sabooni}}, \bibinfo {author} {\bibfnamefont {Q.}~\bibnamefont {Li}},
  \bibinfo {author} {\bibfnamefont {S.}~\bibnamefont {Kr{\"o}ll}},\ and\
  \bibinfo {author} {\bibfnamefont {L.}~\bibnamefont {Rippe}},\ }\bibfield
  {title} {\bibinfo {title} {Efficient quantum memory using a weakly absorbing
  sample},\ }\href@noop {} {\bibfield  {journal} {\bibinfo  {journal} {Physical
  review letters}\ }\textbf {\bibinfo {volume} {110}},\ \bibinfo {pages}
  {133604} (\bibinfo {year} {2013})}\BibitemShut {NoStop}%
\bibitem [{\citenamefont {Laplane}\ \emph {et~al.}(2017)\citenamefont
  {Laplane}, \citenamefont {Jobez}, \citenamefont {Etesse}, \citenamefont
  {Gisin},\ and\ \citenamefont {Afzelius}}]{laplane2017multimode}%
  \BibitemOpen
  \bibfield  {author} {\bibinfo {author} {\bibfnamefont {C.}~\bibnamefont
  {Laplane}}, \bibinfo {author} {\bibfnamefont {P.}~\bibnamefont {Jobez}},
  \bibinfo {author} {\bibfnamefont {J.}~\bibnamefont {Etesse}}, \bibinfo
  {author} {\bibfnamefont {N.}~\bibnamefont {Gisin}},\ and\ \bibinfo {author}
  {\bibfnamefont {M.}~\bibnamefont {Afzelius}},\ }\bibfield  {title} {\bibinfo
  {title} {Multimode and long-lived quantum correlations between photons and
  spins in a crystal},\ }\href@noop {} {\bibfield  {journal} {\bibinfo
  {journal} {Physical review letters}\ }\textbf {\bibinfo {volume} {118}},\
  \bibinfo {pages} {210501} (\bibinfo {year} {2017})}\BibitemShut {NoStop}%
\bibitem [{\citenamefont {Clausen}\ \emph {et~al.}(2011)\citenamefont
  {Clausen}, \citenamefont {Usmani}, \citenamefont {Bussieres}, \citenamefont
  {Sangouard}, \citenamefont {Afzelius}, \citenamefont {de~Riedmatten},\ and\
  \citenamefont {Gisin}}]{clausen2011quantum}%
  \BibitemOpen
  \bibfield  {author} {\bibinfo {author} {\bibfnamefont {C.}~\bibnamefont
  {Clausen}}, \bibinfo {author} {\bibfnamefont {I.}~\bibnamefont {Usmani}},
  \bibinfo {author} {\bibfnamefont {F.}~\bibnamefont {Bussieres}}, \bibinfo
  {author} {\bibfnamefont {N.}~\bibnamefont {Sangouard}}, \bibinfo {author}
  {\bibfnamefont {M.}~\bibnamefont {Afzelius}}, \bibinfo {author}
  {\bibfnamefont {H.}~\bibnamefont {de~Riedmatten}},\ and\ \bibinfo {author}
  {\bibfnamefont {N.}~\bibnamefont {Gisin}},\ }\bibfield  {title} {\bibinfo
  {title} {Quantum storage of photonic entanglement in a crystal},\ }\href@noop
  {} {\bibfield  {journal} {\bibinfo  {journal} {Nature}\ }\textbf {\bibinfo
  {volume} {469}},\ \bibinfo {pages} {508} (\bibinfo {year}
  {2011})}\BibitemShut {NoStop}%
\bibitem [{\citenamefont {Bussi{\`e}res}\ \emph {et~al.}(2014)\citenamefont
  {Bussi{\`e}res}, \citenamefont {Clausen}, \citenamefont {Tiranov},
  \citenamefont {Korzh}, \citenamefont {Verma}, \citenamefont {Nam},
  \citenamefont {Marsili}, \citenamefont {Ferrier}, \citenamefont {Goldner},
  \citenamefont {Herrmann} \emph {et~al.}}]{bussieres2014quantum}%
  \BibitemOpen
  \bibfield  {author} {\bibinfo {author} {\bibfnamefont {F.}~\bibnamefont
  {Bussi{\`e}res}}, \bibinfo {author} {\bibfnamefont {C.}~\bibnamefont
  {Clausen}}, \bibinfo {author} {\bibfnamefont {A.}~\bibnamefont {Tiranov}},
  \bibinfo {author} {\bibfnamefont {B.}~\bibnamefont {Korzh}}, \bibinfo
  {author} {\bibfnamefont {V.~B.}\ \bibnamefont {Verma}}, \bibinfo {author}
  {\bibfnamefont {S.~W.}\ \bibnamefont {Nam}}, \bibinfo {author} {\bibfnamefont
  {F.}~\bibnamefont {Marsili}}, \bibinfo {author} {\bibfnamefont
  {A.}~\bibnamefont {Ferrier}}, \bibinfo {author} {\bibfnamefont
  {P.}~\bibnamefont {Goldner}}, \bibinfo {author} {\bibfnamefont
  {H.}~\bibnamefont {Herrmann}}, \emph {et~al.},\ }\bibfield  {title} {\bibinfo
  {title} {Quantum teleportation from a telecom-wavelength photon to a
  solid-state quantum memory},\ }\href@noop {} {\bibfield  {journal} {\bibinfo
  {journal} {Nature Photonics}\ }\textbf {\bibinfo {volume} {8}},\ \bibinfo
  {pages} {775} (\bibinfo {year} {2014})}\BibitemShut {NoStop}%
\bibitem [{\citenamefont {Maring}\ \emph {et~al.}(2017)\citenamefont {Maring},
  \citenamefont {Farrera}, \citenamefont {Kutluer}, \citenamefont {Mazzera},
  \citenamefont {Heinze},\ and\ \citenamefont
  {de~Riedmatten}}]{maring2017photonic}%
  \BibitemOpen
  \bibfield  {author} {\bibinfo {author} {\bibfnamefont {N.}~\bibnamefont
  {Maring}}, \bibinfo {author} {\bibfnamefont {P.}~\bibnamefont {Farrera}},
  \bibinfo {author} {\bibfnamefont {K.}~\bibnamefont {Kutluer}}, \bibinfo
  {author} {\bibfnamefont {M.}~\bibnamefont {Mazzera}}, \bibinfo {author}
  {\bibfnamefont {G.}~\bibnamefont {Heinze}},\ and\ \bibinfo {author}
  {\bibfnamefont {H.}~\bibnamefont {de~Riedmatten}},\ }\bibfield  {title}
  {\bibinfo {title} {Photonic quantum state transfer between a cold atomic gas
  and a crystal},\ }\href@noop {} {\bibfield  {journal} {\bibinfo  {journal}
  {Nature}\ }\textbf {\bibinfo {volume} {551}},\ \bibinfo {pages} {485}
  (\bibinfo {year} {2017})}\BibitemShut {NoStop}%
\bibitem [{\citenamefont {Bienfait}\ \emph {et~al.}(2016)\citenamefont
  {Bienfait}, \citenamefont {Pla}, \citenamefont {Kubo}, \citenamefont {Zhou},
  \citenamefont {Stern}, \citenamefont {Lo}, \citenamefont {Weis},
  \citenamefont {Schenkel}, \citenamefont {Vion}, \citenamefont {Esteve} \emph
  {et~al.}}]{bienfait2016controlling}%
  \BibitemOpen
  \bibfield  {author} {\bibinfo {author} {\bibfnamefont {A.}~\bibnamefont
  {Bienfait}}, \bibinfo {author} {\bibfnamefont {J.}~\bibnamefont {Pla}},
  \bibinfo {author} {\bibfnamefont {Y.}~\bibnamefont {Kubo}}, \bibinfo {author}
  {\bibfnamefont {X.}~\bibnamefont {Zhou}}, \bibinfo {author} {\bibfnamefont
  {M.}~\bibnamefont {Stern}}, \bibinfo {author} {\bibfnamefont
  {C.}~\bibnamefont {Lo}}, \bibinfo {author} {\bibfnamefont {C.}~\bibnamefont
  {Weis}}, \bibinfo {author} {\bibfnamefont {T.}~\bibnamefont {Schenkel}},
  \bibinfo {author} {\bibfnamefont {D.}~\bibnamefont {Vion}}, \bibinfo {author}
  {\bibfnamefont {D.}~\bibnamefont {Esteve}}, \emph {et~al.},\ }\bibfield
  {title} {\bibinfo {title} {Controlling spin relaxation with a cavity},\
  }\href@noop {} {\bibfield  {journal} {\bibinfo  {journal} {Nature}\ }\textbf
  {\bibinfo {volume} {531}},\ \bibinfo {pages} {74} (\bibinfo {year}
  {2016})}\BibitemShut {NoStop}%
\bibitem [{\citenamefont {Albanese}\ \emph {et~al.}(2020)\citenamefont
  {Albanese}, \citenamefont {Probst}, \citenamefont {Ranjan}, \citenamefont
  {Zollitsch}, \citenamefont {Pechal}, \citenamefont {Wallraff}, \citenamefont
  {Morton}, \citenamefont {Vion}, \citenamefont {Esteve}, \citenamefont
  {Flurin} \emph {et~al.}}]{albanese2020radiative}%
  \BibitemOpen
  \bibfield  {author} {\bibinfo {author} {\bibfnamefont {B.}~\bibnamefont
  {Albanese}}, \bibinfo {author} {\bibfnamefont {S.}~\bibnamefont {Probst}},
  \bibinfo {author} {\bibfnamefont {V.}~\bibnamefont {Ranjan}}, \bibinfo
  {author} {\bibfnamefont {C.~W.}\ \bibnamefont {Zollitsch}}, \bibinfo {author}
  {\bibfnamefont {M.}~\bibnamefont {Pechal}}, \bibinfo {author} {\bibfnamefont
  {A.}~\bibnamefont {Wallraff}}, \bibinfo {author} {\bibfnamefont {J.~J.}\
  \bibnamefont {Morton}}, \bibinfo {author} {\bibfnamefont {D.}~\bibnamefont
  {Vion}}, \bibinfo {author} {\bibfnamefont {D.}~\bibnamefont {Esteve}},
  \bibinfo {author} {\bibfnamefont {E.}~\bibnamefont {Flurin}}, \emph
  {et~al.},\ }\bibfield  {title} {\bibinfo {title} {Radiative cooling of a spin
  ensemble},\ }\href@noop {} {\bibfield  {journal} {\bibinfo  {journal} {Nature
  Physics}\ }\textbf {\bibinfo {volume} {16}},\ \bibinfo {pages} {751}
  (\bibinfo {year} {2020})}\BibitemShut {NoStop}%
\bibitem [{\citenamefont {Williamson}\ \emph {et~al.}(2014)\citenamefont
  {Williamson}, \citenamefont {Chen},\ and\ \citenamefont
  {Longdell}}]{williamson2014magneto}%
  \BibitemOpen
  \bibfield  {author} {\bibinfo {author} {\bibfnamefont {L.~A.}\ \bibnamefont
  {Williamson}}, \bibinfo {author} {\bibfnamefont {Y.-H.}\ \bibnamefont
  {Chen}},\ and\ \bibinfo {author} {\bibfnamefont {J.~J.}\ \bibnamefont
  {Longdell}},\ }\bibfield  {title} {\bibinfo {title} {Magneto-optic modulator
  with unit quantum efficiency},\ }\href@noop {} {\bibfield  {journal}
  {\bibinfo  {journal} {Physical review letters}\ }\textbf {\bibinfo {volume}
  {113}},\ \bibinfo {pages} {203601} (\bibinfo {year} {2014})}\BibitemShut
  {NoStop}%
\bibitem [{\citenamefont {Probst}\ \emph {et~al.}(2014)\citenamefont {Probst},
  \citenamefont {Tkal{\v{c}}ec}, \citenamefont {Rotzinger}, \citenamefont
  {Rieger}, \citenamefont {Le~Floch}, \citenamefont {Goryachev}, \citenamefont
  {Tobar}, \citenamefont {Ustinov},\ and\ \citenamefont
  {Bushev}}]{probst2014three}%
  \BibitemOpen
  \bibfield  {author} {\bibinfo {author} {\bibfnamefont {S.}~\bibnamefont
  {Probst}}, \bibinfo {author} {\bibfnamefont {A.}~\bibnamefont
  {Tkal{\v{c}}ec}}, \bibinfo {author} {\bibfnamefont {H.}~\bibnamefont
  {Rotzinger}}, \bibinfo {author} {\bibfnamefont {D.}~\bibnamefont {Rieger}},
  \bibinfo {author} {\bibfnamefont {J.-M.}\ \bibnamefont {Le~Floch}}, \bibinfo
  {author} {\bibfnamefont {M.}~\bibnamefont {Goryachev}}, \bibinfo {author}
  {\bibfnamefont {M.}~\bibnamefont {Tobar}}, \bibinfo {author} {\bibfnamefont
  {A.}~\bibnamefont {Ustinov}},\ and\ \bibinfo {author} {\bibfnamefont
  {P.}~\bibnamefont {Bushev}},\ }\bibfield  {title} {\bibinfo {title}
  {Three-dimensional cavity quantum electrodynamics with a rare-earth spin
  ensemble},\ }\href@noop {} {\bibfield  {journal} {\bibinfo  {journal}
  {Physical Review B}\ }\textbf {\bibinfo {volume} {90}},\ \bibinfo {pages}
  {100404} (\bibinfo {year} {2014})}\BibitemShut {NoStop}%
\bibitem [{\citenamefont {Fernandez-Gonzalvo}\ \emph
  {et~al.}(2015)\citenamefont {Fernandez-Gonzalvo}, \citenamefont {Chen},
  \citenamefont {Yin}, \citenamefont {Rogge},\ and\ \citenamefont
  {Longdell}}]{fernandez2015coherent}%
  \BibitemOpen
  \bibfield  {author} {\bibinfo {author} {\bibfnamefont {X.}~\bibnamefont
  {Fernandez-Gonzalvo}}, \bibinfo {author} {\bibfnamefont {Y.-H.}\ \bibnamefont
  {Chen}}, \bibinfo {author} {\bibfnamefont {C.}~\bibnamefont {Yin}}, \bibinfo
  {author} {\bibfnamefont {S.}~\bibnamefont {Rogge}},\ and\ \bibinfo {author}
  {\bibfnamefont {J.~J.}\ \bibnamefont {Longdell}},\ }\bibfield  {title}
  {\bibinfo {title} {Coherent frequency up-conversion of microwaves to the
  optical telecommunications band in an \ce{Er}:\ce{YSO} crystal},\ }\href@noop
  {} {\bibfield  {journal} {\bibinfo  {journal} {Physical Review A}\ }\textbf
  {\bibinfo {volume} {92}},\ \bibinfo {pages} {062313} (\bibinfo {year}
  {2015})}\BibitemShut {NoStop}%
\bibitem [{\citenamefont {Chen}\ \emph {et~al.}(2018)\citenamefont {Chen},
  \citenamefont {Fernandez-Gonzalvo}, \citenamefont {Horvath}, \citenamefont
  {Rakonjac},\ and\ \citenamefont {Longdell}}]{chen2018hyperfine}%
  \BibitemOpen
  \bibfield  {author} {\bibinfo {author} {\bibfnamefont {Y.-H.}\ \bibnamefont
  {Chen}}, \bibinfo {author} {\bibfnamefont {X.}~\bibnamefont
  {Fernandez-Gonzalvo}}, \bibinfo {author} {\bibfnamefont {S.~P.}\ \bibnamefont
  {Horvath}}, \bibinfo {author} {\bibfnamefont {J.~V.}\ \bibnamefont
  {Rakonjac}},\ and\ \bibinfo {author} {\bibfnamefont {J.~J.}\ \bibnamefont
  {Longdell}},\ }\bibfield  {title} {\bibinfo {title} {Hyperfine interactions
  of \ce{Er}$^{3+}$ ions in \ce{Y2SiO5}: Electron paramagnetic resonance in a
  tunable microwave cavity},\ }\href@noop {} {\bibfield  {journal} {\bibinfo
  {journal} {Physical Review B}\ }\textbf {\bibinfo {volume} {97}},\ \bibinfo
  {pages} {024419} (\bibinfo {year} {2018})}\BibitemShut {NoStop}%
\bibitem [{\citenamefont {Creedon}\ \emph {et~al.}(2012)\citenamefont
  {Creedon}, \citenamefont {Benmessa{\"\i}}, \citenamefont {Bowen},\ and\
  \citenamefont {Tobar}}]{creedon2012four}%
  \BibitemOpen
  \bibfield  {author} {\bibinfo {author} {\bibfnamefont {D.~L.}\ \bibnamefont
  {Creedon}}, \bibinfo {author} {\bibfnamefont {K.}~\bibnamefont
  {Benmessa{\"\i}}}, \bibinfo {author} {\bibfnamefont {W.~P.}\ \bibnamefont
  {Bowen}},\ and\ \bibinfo {author} {\bibfnamefont {M.~E.}\ \bibnamefont
  {Tobar}},\ }\bibfield  {title} {\bibinfo {title} {Four-wave mixing from
  \ce{Fe}$^{3+}$ spins in sapphire},\ }\href@noop {} {\bibfield  {journal}
  {\bibinfo  {journal} {Physical Review Letters}\ }\textbf {\bibinfo {volume}
  {108}},\ \bibinfo {pages} {093902} (\bibinfo {year} {2012})}\BibitemShut
  {NoStop}%
\bibitem [{\citenamefont {Farr}\ \emph {et~al.}(2015)\citenamefont {Farr},
  \citenamefont {Goryachev}, \citenamefont {Le~Floch}, \citenamefont {Bushev},\
  and\ \citenamefont {Tobar}}]{farr2015evidence}%
  \BibitemOpen
  \bibfield  {author} {\bibinfo {author} {\bibfnamefont {W.~G.}\ \bibnamefont
  {Farr}}, \bibinfo {author} {\bibfnamefont {M.}~\bibnamefont {Goryachev}},
  \bibinfo {author} {\bibfnamefont {J.-M.}\ \bibnamefont {Le~Floch}}, \bibinfo
  {author} {\bibfnamefont {P.}~\bibnamefont {Bushev}},\ and\ \bibinfo {author}
  {\bibfnamefont {M.~E.}\ \bibnamefont {Tobar}},\ }\bibfield  {title} {\bibinfo
  {title} {Evidence of dilute ferromagnetism in rare-earth doped yttrium
  aluminium garnet},\ }\href@noop {} {\bibfield  {journal} {\bibinfo  {journal}
  {Applied Physics Letters}\ }\textbf {\bibinfo {volume} {107}},\ \bibinfo
  {pages} {122401} (\bibinfo {year} {2015})}\BibitemShut {NoStop}%
\bibitem [{\citenamefont {Wisby}\ \emph {et~al.}(2016)\citenamefont {Wisby},
  \citenamefont {De~Graaf}, \citenamefont {Gwilliam}, \citenamefont {Adamyan},
  \citenamefont {Kubatkin}, \citenamefont {Meeson}, \citenamefont
  {Tzalenchuk},\ and\ \citenamefont {Lindstr{\"o}m}}]{wisby2016angle}%
  \BibitemOpen
  \bibfield  {author} {\bibinfo {author} {\bibfnamefont {I.}~\bibnamefont
  {Wisby}}, \bibinfo {author} {\bibfnamefont {S.}~\bibnamefont {De~Graaf}},
  \bibinfo {author} {\bibfnamefont {R.}~\bibnamefont {Gwilliam}}, \bibinfo
  {author} {\bibfnamefont {A.}~\bibnamefont {Adamyan}}, \bibinfo {author}
  {\bibfnamefont {S.}~\bibnamefont {Kubatkin}}, \bibinfo {author}
  {\bibfnamefont {P.}~\bibnamefont {Meeson}}, \bibinfo {author} {\bibfnamefont
  {A.~Y.}\ \bibnamefont {Tzalenchuk}},\ and\ \bibinfo {author} {\bibfnamefont
  {T.}~\bibnamefont {Lindstr{\"o}m}},\ }\bibfield  {title} {\bibinfo {title}
  {Angle-dependent microresonator esr characterization of locally doped
  \ce{Gd}$^{3+}$:\ce{Al2O3}},\ }\href@noop {} {\bibfield  {journal} {\bibinfo
  {journal} {Physical Review Applied}\ }\textbf {\bibinfo {volume} {6}},\
  \bibinfo {pages} {024021} (\bibinfo {year} {2016})}\BibitemShut {NoStop}%
\bibitem [{\citenamefont {Bushev}\ \emph {et~al.}(2011)\citenamefont {Bushev},
  \citenamefont {Feofanov}, \citenamefont {Rotzinger}, \citenamefont
  {Protopopov}, \citenamefont {Cole}, \citenamefont {Wilson}, \citenamefont
  {Fischer}, \citenamefont {Lukashenko},\ and\ \citenamefont
  {Ustinov}}]{bushev2011ultralow}%
  \BibitemOpen
  \bibfield  {author} {\bibinfo {author} {\bibfnamefont {P.}~\bibnamefont
  {Bushev}}, \bibinfo {author} {\bibfnamefont {A.}~\bibnamefont {Feofanov}},
  \bibinfo {author} {\bibfnamefont {H.}~\bibnamefont {Rotzinger}}, \bibinfo
  {author} {\bibfnamefont {I.}~\bibnamefont {Protopopov}}, \bibinfo {author}
  {\bibfnamefont {J.}~\bibnamefont {Cole}}, \bibinfo {author} {\bibfnamefont
  {C.}~\bibnamefont {Wilson}}, \bibinfo {author} {\bibfnamefont
  {G.}~\bibnamefont {Fischer}}, \bibinfo {author} {\bibfnamefont
  {A.}~\bibnamefont {Lukashenko}},\ and\ \bibinfo {author} {\bibfnamefont
  {A.}~\bibnamefont {Ustinov}},\ }\bibfield  {title} {\bibinfo {title}
  {Ultralow-power spectroscopy of a rare-earth spin ensemble using a
  superconducting resonator},\ }\href@noop {} {\bibfield  {journal} {\bibinfo
  {journal} {Physical Review B}\ }\textbf {\bibinfo {volume} {84}},\ \bibinfo
  {pages} {060501} (\bibinfo {year} {2011})}\BibitemShut {NoStop}%
\bibitem [{\citenamefont {Staudt}\ \emph {et~al.}(2012)\citenamefont {Staudt},
  \citenamefont {Hoi}, \citenamefont {Krantz}, \citenamefont {Sandberg},
  \citenamefont {Simoen}, \citenamefont {Bushev}, \citenamefont {Sangouard},
  \citenamefont {Afzelius}, \citenamefont {Shumeiko}, \citenamefont {Johansson}
  \emph {et~al.}}]{staudt2012coupling}%
  \BibitemOpen
  \bibfield  {author} {\bibinfo {author} {\bibfnamefont {M.~U.}\ \bibnamefont
  {Staudt}}, \bibinfo {author} {\bibfnamefont {I.-C.}\ \bibnamefont {Hoi}},
  \bibinfo {author} {\bibfnamefont {P.}~\bibnamefont {Krantz}}, \bibinfo
  {author} {\bibfnamefont {M.}~\bibnamefont {Sandberg}}, \bibinfo {author}
  {\bibfnamefont {M.}~\bibnamefont {Simoen}}, \bibinfo {author} {\bibfnamefont
  {P.}~\bibnamefont {Bushev}}, \bibinfo {author} {\bibfnamefont
  {N.}~\bibnamefont {Sangouard}}, \bibinfo {author} {\bibfnamefont
  {M.}~\bibnamefont {Afzelius}}, \bibinfo {author} {\bibfnamefont {V.~S.}\
  \bibnamefont {Shumeiko}}, \bibinfo {author} {\bibfnamefont {G.}~\bibnamefont
  {Johansson}}, \emph {et~al.},\ }\bibfield  {title} {\bibinfo {title}
  {Coupling of an erbium spin ensemble to a superconducting resonator},\
  }\href@noop {} {\bibfield  {journal} {\bibinfo  {journal} {Journal of Physics
  B: Atomic, Molecular and Optical Physics}\ }\textbf {\bibinfo {volume}
  {45}},\ \bibinfo {pages} {124019} (\bibinfo {year} {2012})}\BibitemShut
  {NoStop}%
\bibitem [{\citenamefont {Probst}\ \emph {et~al.}(2013)\citenamefont {Probst},
  \citenamefont {Rotzinger}, \citenamefont {W{\"u}nsch}, \citenamefont {Jung},
  \citenamefont {Jerger}, \citenamefont {Siegel}, \citenamefont {Ustinov},\
  and\ \citenamefont {Bushev}}]{probst2013anisotropic}%
  \BibitemOpen
  \bibfield  {author} {\bibinfo {author} {\bibfnamefont {S.}~\bibnamefont
  {Probst}}, \bibinfo {author} {\bibfnamefont {H.}~\bibnamefont {Rotzinger}},
  \bibinfo {author} {\bibfnamefont {S.}~\bibnamefont {W{\"u}nsch}}, \bibinfo
  {author} {\bibfnamefont {P.}~\bibnamefont {Jung}}, \bibinfo {author}
  {\bibfnamefont {M.}~\bibnamefont {Jerger}}, \bibinfo {author} {\bibfnamefont
  {M.}~\bibnamefont {Siegel}}, \bibinfo {author} {\bibfnamefont
  {A.}~\bibnamefont {Ustinov}},\ and\ \bibinfo {author} {\bibfnamefont
  {P.}~\bibnamefont {Bushev}},\ }\bibfield  {title} {\bibinfo {title}
  {Anisotropic rare-earth spin ensemble strongly coupled to a superconducting
  resonator},\ }\href@noop {} {\bibfield  {journal} {\bibinfo  {journal}
  {Physical Review Letters}\ }\textbf {\bibinfo {volume} {110}},\ \bibinfo
  {pages} {157001} (\bibinfo {year} {2013})}\BibitemShut {NoStop}%
\bibitem [{\citenamefont {Tkal{\v{c}}ec}\ \emph {et~al.}(2014)\citenamefont
  {Tkal{\v{c}}ec}, \citenamefont {Probst}, \citenamefont {Rieger},
  \citenamefont {Rotzinger}, \citenamefont {W{\"u}nsch}, \citenamefont
  {Kukharchyk}, \citenamefont {Wieck}, \citenamefont {Siegel}, \citenamefont
  {Ustinov},\ and\ \citenamefont {Bushev}}]{tkalvcec2014strong}%
  \BibitemOpen
  \bibfield  {author} {\bibinfo {author} {\bibfnamefont {A.}~\bibnamefont
  {Tkal{\v{c}}ec}}, \bibinfo {author} {\bibfnamefont {S.}~\bibnamefont
  {Probst}}, \bibinfo {author} {\bibfnamefont {D.}~\bibnamefont {Rieger}},
  \bibinfo {author} {\bibfnamefont {H.}~\bibnamefont {Rotzinger}}, \bibinfo
  {author} {\bibfnamefont {S.}~\bibnamefont {W{\"u}nsch}}, \bibinfo {author}
  {\bibfnamefont {N.}~\bibnamefont {Kukharchyk}}, \bibinfo {author}
  {\bibfnamefont {A.}~\bibnamefont {Wieck}}, \bibinfo {author} {\bibfnamefont
  {M.}~\bibnamefont {Siegel}}, \bibinfo {author} {\bibfnamefont
  {A.}~\bibnamefont {Ustinov}},\ and\ \bibinfo {author} {\bibfnamefont
  {P.}~\bibnamefont {Bushev}},\ }\bibfield  {title} {\bibinfo {title} {Strong
  coupling of an \ce{Er}$^{3+}$-doped \ce{YAlO3} crystal to a superconducting
  resonator},\ }\href@noop {} {\bibfield  {journal} {\bibinfo  {journal}
  {Physical Review B}\ }\textbf {\bibinfo {volume} {90}},\ \bibinfo {pages}
  {075112} (\bibinfo {year} {2014})}\BibitemShut {NoStop}%
\bibitem [{\citenamefont {Dold}\ \emph {et~al.}(2019)\citenamefont {Dold},
  \citenamefont {Zollitsch}, \citenamefont {O’sullivan}, \citenamefont
  {Welinski}, \citenamefont {Ferrier}, \citenamefont {Goldner}, \citenamefont
  {de~Graaf}, \citenamefont {Lindstr{\"o}m},\ and\ \citenamefont
  {Morton}}]{dold2019high}%
  \BibitemOpen
  \bibfield  {author} {\bibinfo {author} {\bibfnamefont {G.}~\bibnamefont
  {Dold}}, \bibinfo {author} {\bibfnamefont {C.~W.}\ \bibnamefont {Zollitsch}},
  \bibinfo {author} {\bibfnamefont {J.}~\bibnamefont {O’sullivan}}, \bibinfo
  {author} {\bibfnamefont {S.}~\bibnamefont {Welinski}}, \bibinfo {author}
  {\bibfnamefont {A.}~\bibnamefont {Ferrier}}, \bibinfo {author} {\bibfnamefont
  {P.}~\bibnamefont {Goldner}}, \bibinfo {author} {\bibfnamefont
  {S.}~\bibnamefont {de~Graaf}}, \bibinfo {author} {\bibfnamefont
  {T.}~\bibnamefont {Lindstr{\"o}m}},\ and\ \bibinfo {author} {\bibfnamefont
  {J.~J.}\ \bibnamefont {Morton}},\ }\bibfield  {title} {\bibinfo {title}
  {High-cooperativity coupling of a rare-earth spin ensemble to a
  superconducting resonator using yttrium orthosilicate as a substrate},\
  }\href@noop {} {\bibfield  {journal} {\bibinfo  {journal} {Physical Review
  Applied}\ }\textbf {\bibinfo {volume} {11}},\ \bibinfo {pages} {054082}
  (\bibinfo {year} {2019})}\BibitemShut {NoStop}%
\bibitem [{\citenamefont {Guillot-No{\"e}l}\ \emph {et~al.}(2006)\citenamefont
  {Guillot-No{\"e}l}, \citenamefont {Goldner}, \citenamefont {Le~Du},
  \citenamefont {Baldit}, \citenamefont {Monnier},\ and\ \citenamefont
  {Bencheikh}}]{guillot2006hyperfine}%
  \BibitemOpen
  \bibfield  {author} {\bibinfo {author} {\bibfnamefont {O.}~\bibnamefont
  {Guillot-No{\"e}l}}, \bibinfo {author} {\bibfnamefont {P.}~\bibnamefont
  {Goldner}}, \bibinfo {author} {\bibfnamefont {Y.}~\bibnamefont {Le~Du}},
  \bibinfo {author} {\bibfnamefont {E.}~\bibnamefont {Baldit}}, \bibinfo
  {author} {\bibfnamefont {P.}~\bibnamefont {Monnier}},\ and\ \bibinfo {author}
  {\bibfnamefont {K.}~\bibnamefont {Bencheikh}},\ }\bibfield  {title} {\bibinfo
  {title} {Hyperfine interaction of \ce{Er}$^{3+}$ ions in \ce{Y2SiO5}: An
  electron paramagnetic resonance spectroscopy study},\ }\href@noop {}
  {\bibfield  {journal} {\bibinfo  {journal} {Physical Review B}\ }\textbf
  {\bibinfo {volume} {74}},\ \bibinfo {pages} {214409} (\bibinfo {year}
  {2006})}\BibitemShut {NoStop}%
\bibitem [{\citenamefont {Sun}\ \emph {et~al.}(2008)\citenamefont {Sun},
  \citenamefont {B{\"o}ttger}, \citenamefont {Thiel},\ and\ \citenamefont
  {Cone}}]{sun2008magnetic}%
  \BibitemOpen
  \bibfield  {author} {\bibinfo {author} {\bibfnamefont {Y.}~\bibnamefont
  {Sun}}, \bibinfo {author} {\bibfnamefont {T.}~\bibnamefont {B{\"o}ttger}},
  \bibinfo {author} {\bibfnamefont {C.}~\bibnamefont {Thiel}},\ and\ \bibinfo
  {author} {\bibfnamefont {R.}~\bibnamefont {Cone}},\ }\bibfield  {title}
  {\bibinfo {title} {Magnetic g tensors for the $^4$i$_{15/2}$ and
  $^4$i$_{13/2}$ states of \ce{Er}$^{3+}$: \ce{Y2SiO5}},\ }\href@noop {}
  {\bibfield  {journal} {\bibinfo  {journal} {Physical Review B}\ }\textbf
  {\bibinfo {volume} {77}},\ \bibinfo {pages} {085124} (\bibinfo {year}
  {2008})}\BibitemShut {NoStop}%
\bibitem [{\citenamefont {Stoll}\ and\ \citenamefont
  {Schweiger}(2006)}]{stoll2006easyspin}%
  \BibitemOpen
  \bibfield  {author} {\bibinfo {author} {\bibfnamefont {S.}~\bibnamefont
  {Stoll}}\ and\ \bibinfo {author} {\bibfnamefont {A.}~\bibnamefont
  {Schweiger}},\ }\bibfield  {title} {\bibinfo {title} {Easyspin, a
  comprehensive software package for spectral simulation and analysis in epr},\
  }\href@noop {} {\bibfield  {journal} {\bibinfo  {journal} {Journal of
  magnetic resonance}\ }\textbf {\bibinfo {volume} {178}},\ \bibinfo {pages}
  {42} (\bibinfo {year} {2006})}\BibitemShut {NoStop}%
\end{thebibliography}%

\end{document}